# The Energetic Cost of Building a Virus

Gita Mahmoudabadi[1], Ron Milo[2], Rob Phillips[1,3]

*1 Department of Bioengineering, California Institute of Technology, Pasadena, CA 91125, USA. 2 Department of Plant and Environmental Sciences, Weizmann Institute of Science, Rehovot 7610001, Israel. 3 Department of Applied Physics, California Institute of Technology, Pasadena, CA 91125, USA.*

## Abstract

Viruses are incapable of autonomous energy production. Although many experimental studies make it clear that viruses are parasitic entities that hijack the host's molecular resources, a detailed estimate for the energetic cost of viral synthesis is largely lacking. To quantify the energetic cost of viruses to their hosts, we enumerated the costs associated with two very distinct but representative DNA and RNA viruses, namely, T4 and influenza. We found that for these viruses, translation of viral proteins is the most energetically expensive process. Interestingly, the cost of building a T4 phage and a single influenza virus are nearly the same. Due to influenza's higher burst size, however, the overall cost of a T4 phage infection is only 2 - 3% of the cost of an influenza infection. The costs of these infections relative to their host's estimated energy budget during the infection reveal that a T4 infection consumes about a third of its host's energy budget, whereas an influenza infection consumes only 1%. Building on our estimates for T4, we show how the energetic costs of double-stranded DNA viruses scale with virus size, revealing that the dominant cost of building a virus can switch from translation to genome replication above a critical virus size. Lastly, using our predictions for the energetic cost of viruses, we provide estimates for the strengths of selection and genetic drift acting on newly incorporated genetic elements in viral genomes, under conditions of energy limitation.

## Significance Statement

Viruses rely entirely on their host as an energy source. Despite numerous experimental studies that demonstrate the capability of viruses to rewire and undermine their host's metabolism, we still largely lack a quantitative understanding of an infection's energetics. And yet, the energetics of a viral infection is at the center of broader evolutionary and physical questions in virology. By enumerating the energetic costs of different viral processes, we open the door to quantitative predictions about viral evolution. For example, we predict that for the majority of viruses, translation will serve as the dominant cost of building a virus, and that selection, rather than drift, will govern the fate of new genetic elements within viral genomes.

## Introduction

Viruses are biological ‘entities’ at the boundary of life. Without cells to infect, viruses as we know them would cease to function, as they rely on their hosts to replicate. Though the extent of this reliance varies for different viruses, all viruses consume from the host’s energy budget in creating the next generation of viruses. There are many examples of viruses that actively subvert the host transcriptional and translational processes in favor of their own replication (1). This viral takeover of the host metabolism manifests itself in a variety of forms such as in the degradation of the host’s genome or the inhibition of the host’s mRNA translation (1). These examples suggest that a viral infection requires a considerable amount of the host’s energetic supply. In support of this view are experiments on T4 (2), T7 (3), *Pseudoalteromonas* phage (4), and *Paramecium bursaria chlorella virus-1* or PBCV-1 (5), demonstrating the viral burst size to correlate positively with the host growth rate. In the case of PBCV-1, the burst size is reduced by 50% when its photosynthetic host, a freshwater algae, is grown in the dark (5). Similarly, slow growing *E. coli* with a doubling time of 21 hours affords a T4 burst size of just one phage (6), as opposed to a burst size of 100-200 phages during optimal growth conditions.

There are many other experimental studies (discussed in the SI section I) (7-11) that demonstrate viruses to be capable of rewiring the host metabolism. These fascinating observations led us to ask the following questions: what is the energetic cost of a viral infection, and what is the energetic burden of a viral infection on the host cell? To our knowledge, the first attempt to address these problems is provided through a kinetic model of the growth of Qß phage, which demonstrates that Qß growth is energetically optimal (12). A more recent study performed numerical simulations of the impact of a phage T7 infection on its *E. coli* host, yielding very interesting insights into the time course of the metabolic demands of a viral infection (13).

To further explore the energetic requirements of viral synthesis, we made careful estimates of the energetic costs for two viruses with very different characteristics, namely the T4 phage and the influenza A virus. T4 phage is a double-stranded DNA (dsDNA) virus with a 169 kb genome that infects *E. coli*. The influenza virus is a negative-sense, single-stranded RNA virus (-ssRNA) with a segmented genome that is 10.6 kb in total length. The influenza virus is a eukaryotic virus infecting various animals, with an average burst size of 6000 (14). Similar to many other dsDNA viruses, T4 phage infections yield a relatively modest burst size, with the majority of T4 phages resulting in a burst size of approximately 200 (15). To determine the energetic demand of viruses on their hosts, the cost estimate for building a single virus has to be multiplied by the viral burst size and placed in the context of the host's energy budget during the viral infection.

Concretely, the costs associated with building a virus can be broken down into the following processes that are common to the life-cycles of many viruses: 1) viral entry 2) intracellular transport, 3) genome replication,4) transcription, 5) translation, 6) assembly and genome packaging, and 7) exit. Detailed estimates for all of these costs are provided in the SI. Our strategy was to examine each of these processes for both viruses in parallel, comparing and contrasting the energetic burdens of each of the steps in the viral life-cycle.

## Results

By estimating the energetic costs of influenza and T4 life-cycles, we show that surprisingly the cost of synthesizing an influenza virus and a T4 phage are nearly the same (Table 1). The outcome of the analysis to be discussed in the remainder of the paper is summarized pictorially in Figure 1 for bacteriophage T4 and Figure 2 for influenza. For both viruses, the energetic cost of translation outweighs other costs (Table 1, Figures 1, 2, 3), though as we will show at the end of the paper, since translation scales with the surface area of the viral capsid and replication scales as the volume of the virus, for double-stranded DNA phages larger than a critical size, the replication cost outpaces the translation cost. Our results will be provided in terms of two different energetic cost definitions described in detail as part of SI sections II-IV.

To briefly summarize, in our first definition of energetic cost, termed direct cost or $E_D$, we will only account for hydrolysis of ATP molecules (and equivalent molecules, such as GTP) required during viral synthesis. This definition will include costs such as those incurred during the synthesis and polymerization of building blocks (SI Figure 1A, steps 3 and 4; SI section II). In our second definition, termed total cost or $E_T$, we not only account for the direct costs, but also for the opportunity cost of building blocks, $E_O$, required during viral synthesis (SI Figure 1A, steps 1-4; SI sections II-IV); hence, $E_T = E_D + E_O$. We define the opportunity cost of a building block as the number of ATP molecules that *could have been generated had the building block not been synthesized*. The full definition and derivation of these two cost components can be found in the SI (SI sections II-IV, SI table 5, SI Figure 1, SI Figure 2).

The distinction between these two different energetic cost definitions is that under the direct cost definition, we attribute energy only to the hydrolysis of ATP-equivalent molecules, whereas under the total cost definition, we also attribute an energetic cost to the building blocks that are usurped from the host during viral synthesis. Both energy definitions have physical significance. For example, the direct cost definition only accounts for ATP (and ATP-equivalent) hydrolysis events, thereby giving us insight into heat production and power consumption of a viral infection. The total cost definition, on the other hand, is aligned with traditional energetic cost estimates made from growth experiments in chemostats and allows for a clear comparison between the cost of an infection and the cost of a cell. To help the reader discern between opportunity and direct costs, we will signify the former in units of ^P and the latter in the units of ~P. When reporting total cost estimates, we will simply use P to signify the sum of opportunity and direct costs.

Moreover, in formulating our estimates, we will generally estimate the cost of a certain viral process for a single virus, and then multiply this cost by the viral burst size to determine the infection cost of a given process. Subscript $v$ will denote the cost estimates made for a single virus, and the subscript $i$ will refer to a cost estimate made for an infection. We relegate the energetic cost estimates for all viral process to the SI sections V-XI.

**The direct, opportunity and total costs of T4 and Influenza.** To get a sense for the numbers, here we provide order-of-magnitude estimates of both the costs of translation and replication and refer the interested reader to the SI sections II-XI for full details. As detailed in the SI Tables 1 and 2, both T4 and influenza are comprised of about $10^6$ amino acids. We can estimate the total cost of translation by appealing to a few simple facts. First, the average opportunity cost per amino acid is about 30 ^P. Second, the average direct cost to produce amino acids from precursor metabolites is 2 ~P. Finally, each polypeptide bond incurs a direct cost of 4 ~P. We can see that the total cost of an amino acid is approximately 36 P (30 ^P + 6 ~P). As a result, the translational cost of an influenza virus and a T4 phage both fall between $10^7$ to $10^8$ P (Table 1).

The cost of viral replication can be approximated in a similar fashion: we have to consider that the T4 genome is comprised of roughly 4 x$10^5$ DNA bases and that the influenza genome is composed of an order of magnitude fewer RNA bases ($\approx 10^4$). The total costs of a DNA nucleotide and an RNA nucleotide, including the opportunity costs as well as the direct costs of synthesis and polymerization, are approximately 50 P (SI section II-IX, SI Figure 1, SI Figure 2, SI Table 5). As a result of T4's longer genome length, its total cost of replication ($\approx 10^7$ P) is about an order of magnitude higher than that of an influenza genome (Table 1, Figure 1, Figure 2, SI section VII).

The direct, opportunity and total cost estimates of different viral processes during T4 and influenza infections are summarized in Figures 1-3 and Table 1. The overall cost of a T4 infection is obtained by summing the costs of replication ($E_{REP/i}$), transcription ($E_{TX/i}$), translation ($E_{TL/i}$), and genome packaging ($E_{Pack/i}$) required during the infection (SI sections V-XI, Table 1, Figure 1, Figure 3). These costs together amount to ≈3 x $10^9$ ~P, 8 x $10^9$ ^P, and 1 x $10^{10}$ P, respectively (SI sections V-XI, Table 1, Figure 1, Figure 3). The total cost of a T4 infection is also equivalent to the aerobic respiration of ≈4 x $10^8$ glucose molecules by *E. coli* (26 P per glucose, (16)). Alternatively, it is equivalent to ≈2 x $10^{11}$ $k_BT$ (assuming 1 ATP = 20 $k_BT$ on average) (17).

Similarly, the cost of an influenza infection is obtained by adding up the costs of entry ($E_{Entry}$), intracellular transport ($E_{Transit/i}$), replication ($E_{REP/i}$), transcription ($E_{TX/i}$), translation ($E_{TL/i}$), and exit ($E_{Exit/i}$) required during the infection (SI sections V-XI, Table 1, Figure 2, Figure 3). These processes have a cumulative cost of ≈8 x $10^{10}$ ~P, 5 x $10^{11}$ ^P and 6 x $10^{11}$ P, respectively. The sum of costs in an influenza infection is equivalent to the aerobic respiration of ≈2 x $10^{10}$ glucose molecules by a eukaryotic cell (32 P per glucose). It is also equivalent to ≈$10^{13}$ $k_BT$. It is interesting to note that for both viral infections the opportunity cost components are the dominant component of the total costs.

The direct cost of a T4 phage infection is therefore only ≈3% of the direct cost of an influenza infection. Similarly, the total cost of a T4 phage infection is only ≈2% of the total cost of an influenza infection even though individually a T4 phage and an influenza virus have comparable energetic costs. To contextualize these numbers, the host energy budget (or the host energetic cost, depending on the viewpoint of a virus versus a cell) during the infection has to be taken into account.

The total cost of a cell is experimentally tractable through growth experiments in chemostats, in which cultures are maintained at a constant growth rate. The number of glucose molecules taken up per cell per unit time can be determined. The number of glucose molecules can then be converted to an energetic supply by assuming typical conversion rates of 26 P or 32 P per glucose molecule depending on the organism (16). This energetic cost estimate will be a total cost estimate because not all glucose molecules taken up by the cell are fully metabolized to carbon dioxide and water to generate ATPs. During the cellular life-cycle, the cell has to double its number of building blocks prior to division, and to do so, a fraction of glucose molecules taken up is diverted away from energy production towards biosynthesis pathways. Hence, cellular energetic cost estimates that are derived from chemostat experiments are total cost estimates because they report on the combined opportunity and direct costs of a cell (SI section II-IV).

Based on chemostat growth experiments (18), the total cost of a bacterium and a mammalian cell with volumes of 1 $\mu m^3$ and 2000 $\mu m^3$, respectively, are ≈3 x $10^{10}$ P and ≈5 x $10^{13}$ P, during the course of their viral infections (SI section XII). A simpler estimate for arriving at the total cost of *E. coli* during its 30-minute doubling time is by considering the dry weight of *E. coli* (≈0.6 pg) ((19), BNID 100089). Given that about half of the cell's dry weight is comprised of carbon ((19), BNID 100649), an *E. coli* is composed of ≈2 x $10^{10}$ carbons, supplied from ≈3 x $10^9$ glucose molecules, since each glucose contributes 6 carbons. With the 26 P per glucose conversion for *E. coli*, this is equivalent to a total cost of ≈7 x $10^{10}$ P, which is similar to the number obtained from chemostat growth experiments (18)(SI section XII).

Moreover, we estimate the fractional cost of a viral infection as the ratio of total cost of an infection, $E_{T/i}$, to the total cost of the host during the infection, $E_{T/h}$. For the T4 infection with a burst size of 200 virions, $E_{T/i}$ ≈1 x $10^{10}$ P (Table 1) and $E_{T/h}$ ≈3 x $10^{10}$ P, therefore the fractional cost of the T4 infection is ≈0.3. Interestingly, a calorimetric study of a marine microbial community demonstrated that 25% of the heat released by microbes is due to phage activity (20) – an observation that resonates well with our estimate. In contrast, the influenza infection despite its larger burst size (6000 virions) and higher $E_{T/i}$ (≈6 x $10^{11}$ P) has a fractional cost of just 0.01.

In our estimates for heat production and power consumption of a viral infection, we will not include the total cost of an infection as it contains the opportunity costs; by definition, these opportunity costs do not represent direct expenditure of ATP-equivalent molecules and therefore do not substantially contribute to heat production. In contrast, direct cost estimates capture only the number of ATP-equivalent molecules hydrolyzed during an infection (SI section II).

T4 infection has a direct cost of $3 \times 10^9$ ~P (Table 1). Assuming ATP hydrolysis generates -30 kJ/mole, the heat generated during a T4 infection is approximately 0.1 nJ. An influenza infection with a direct cost of $7 \times 10^{10}$ ~P generates 4 nJ. While influenza infection results in an order of magnitude more heat, the average rate of heat production or the power of T4 and influenza infections are surprisingly very similar. In half an hour, the T4 infection results in the hydrolysis of ATP-equivalent molecules at an average rate of $1 \times 10^6$ ~P per second. In half a day, an influenza infection has an average ATP-hydrolysis rate of $2 \times 10^6$ ~P per second, which is nearly the same rate as that of a T4 infection. Put in terms of the more familiar units of Watts, the power of both viral infections is on the order of 100 fW.

**Generalizing viral energetics for double-stranded DNA viruses.** While we have concluded that for the influenza virus and the T4 phage the translational cost outweighs the replication cost, the ratio of these two costs varies according to the dimensions of a virus. In the case of T4 and influenza, these two viruses had comparable dimensions and consequently were comprised of a similar number of amino acids (SI Tables 1 and 2). However, due to the diminishing surface area to volume ratio of a spherical object as it grows in size, the ratio of translational cost to replication cost also diminishes with increasing radius of a spherical capsid. This simple rule governs not just nucleotide or amino acid composition of a virus, but more fundamentally, it governs the elemental composition of viruses with spherical-like geometries (21).

The full derivation of replication and translational cost estimates as a function of viral capsid inner radius, $r$, can be found in the SI section XIII. From these expressions, it is clear that the translational cost of a virus scales with $r^2$, whereas the replication cost scales with $r^3$ (Figure 4). The critical radius at which replication will outweigh translation in cost is 59 nm for total cost estimates, $r_{crit-Tot}$ (Figure 4, SI section XIII). For the direct cost estimates, the critical radius, $r_{crit-Dir}$, is 42 nm. Interestingly, a survey of structural diversity encompassing 2,600 viruses inhabiting the world's oceans reveals that the average outer capsid radius is 28 nm (22), which is much smaller than the predicted critical radii (Figure 4). As such, for the majority of viruses, we predict translation is the dominant cost of a viral infection.

Furthermore, we provide genome replication to translation cost ratios for about 30 different double-stranded phages (SI Table 3, Figure 4). While we have omitted calculations for the virus tails, they can be simply treated as hollow cylinders and will further decrease the expected replication to translation cost ratio for the tailed viruses. Although we have calculated these ratios for this select group of viruses, similar principles can be applied to modeling the energetics of other viral groups.

**Forces of evolution operating on viral genomes**. Inspired by efforts to consider the evolutionary implications of the cost of a gene to cells of different sizes (18, 23), we were curious whether similar considerations might be in play in the context of viruses. For example, we asked which evolutionary forces are prominently operating on neutral genetic elements that are incorporated into viral genomes, either by horizontal gene transfer, gene duplication or other similar types of events. We further asked whether the viral size is a parameter of interest in the tug of war between different forces of evolution. We will address these topics by assuming that the host lives in an energy-limited environment and that the viral infection, consistent with our findings for T4, consumes a substantial portion of the host energy budget. By making these assumptions, we are able to treat the energetic cost of a genetic element as a fitness cost.

For a genetic element to remain in the population, regardless of whether it is beneficial or not, it must face the consequences of genetic drift which scales with the viral effective population size, $N_e$, as ${N_e}^{-1}$. We follow the treatment of Lynch and Marinov who argue that the net selective advantage of a genetic element is $s_n = s_a - s_c$, where $s_a$ and $s_c$ denote the selective advantage and disadvantage, respectively (Figure 5B). For a genetic element within a viral genome that is non-transcribed and non-translated (Figure 5C), only the energetic cost of its replication poses a selective disadvantage. Assuming the genetic element provides no benefit to the virus ($s_a = 0$), the net selective advantage can be stated as $s_n = -s_c$, the absolute value of which must be much greater than ${N_e}^{-1}$ for selection to operate effectively. Following Lynch and Marinov and others (23, 24), we make the simplifying assumption that a neutral genetic element's selection coefficient, $s_c$, is proportional to its fractional energetic cost, $E_g$ (Figure 5C). In the case of a non-transcribed genetic element, $E_g = \frac{E_{REP/v}}{E_v}$, where $E_{REP/v}$ corresponds to its replication cost and $E_v$ is the sum of all costs of a virus (Figure 5C).

Given that replication cost scales as $r^3$ the effects of selection relative to genetic drift could be different for viruses of different sizes. Consider Virus A, having a radius that is two times larger than that of Virus B (Figure 5D). Because both viruses are assumed to have radii larger than the critical radius, we imagine the scenario in which the cost of genome replication is the dominant cost of synthesizing these viruses. The fractional cost of a genetic element in the smaller virus, $E_{g_Virus\ B}$ is then equal to $8E_{g_Virus\ A}$, where $E_{g_Virus\ A}$ is the fractional cost of the genetic element in the larger virus. This is because the length of the genome is proportional to $r^3$, and consequently, $E_g$ is inversely proportional to $r^3$ (Figure 5D).

Figure 5E and SI Table 4 provide $E_g$ estimates for genetic elements of different lengths (1 – 10,000 base pairs) within 30 dsDNA viruses. To illustrate the effect of scaling in the example provided above, we made the simplifying assumption that the viruses are large enough that their $E_v$ are approximately equal to their replication costs. However, for $E_v$ values in Figure 5E and SI Table 4, we provide more precise estimates, treating $E_v$ as the sum of both the replication cost and the translational cost of a virus. The cost of replicating a double-stranded genetic element can be obtained from SI Eq. 3. For a 1 kb element, which is about the average length of a bacterial gene, the direct and total costs of its replication per virus, $E_{REP/v}$, are 3 x $10^4$ ~P and 9 x $10^4$ P, respectively. Both direct and total cost estimates indicate that the strength of selection acting on a 1 kb, non-transcribed element ranges from 2 x $10^{-2}$ - 7 x $10^{-6}$ (SI Table 4, Figure 5E) when considering viruses with radii ranging from ~20 nm to 400 nm. The difference between direct and total estimates of selection strength is minimal within this range of capsid radii and continues to diminish as the capsids grow in size.

To examine whether selection or genetic drift will decide the fate of a genetic element we need to assess each virus's effective population size. This is difficult because the effective population size of most viruses is unknown and subject to great variability due to several environmental factors (25). The current effective population size estimates regarding HIV, influenza, dengue, and measles fall within $10^1$ to $10^5$ (25-27). Based on the wide range of variation in these effective population sizes, it is difficult to make conclusive statements. It is, however, apparent that the strength of selection on neutral genetic elements is a non-linear function of the viral capsid radius and becomes much weaker as viruses get larger (Figure 5E). In fact, for giant viruses (with outer radius, R > 200 nm), assuming an ${N_e}^{-1}$ = $10^{-5}$, genetic drift could overpower selection, allowing for the persistence of neutral elements of lengths 100 bp or shorter in the population. For the majority of viruses (R = 28 $\pm$ 6.5 nm, (22)), however, selection is likely to be the dominant force and drift may only play a role for genetic elements that are just a few base pairs long (Figure 5E, SI Table 4).

## Discussion

There have been several experiments that imply a viral infection requires a significant portion of the host energy budget (5, 6, 8, 10, 28-30). Following these experimental hints, we enumerated the energetic requirements of two very different viruses on the basis of their life-cycles, and thereby estimated the energetic burdens of these viral infections on the host cells. According to our total cost estimates, a T4 infection with a burst size of 200 will consume a significant portion (about 30%) of the host energy supply. This result, demonstrating a significant fraction of the host energy used by an infection, supports the experimental findings that the T4 burst size is correlated positively with the host growth rate (2, 6). It also lends further credence to the hypothesis that auxiliary metabolic genes within phage genomes are not just evolutionary accidents; rather, they have come to serve a functional role in boosting the host's metabolic capacity, which translates into larger viral burst sizes (8, 9, 30, 31). These calculations make it all the more interesting to develop high-precision, single-cell calorimetry techniques to monitor energy usage during viral infections. Perhaps the most promising support for T4's cost estimate is the observation that the maximum T4 burst size is 1,000 virions (15). Using the total cost to make new viruses, at a burst size of 1,000, the viral infection would consume 170% of the host energy supply, consistent with the observed apparent upper limits on burst size.

While there are several fascinating studies that explore the link between the host metabolism and phage infections (8, 11-13), similar studies focusing on viruses of multicellular eukaryotes are largely lacking. To that end, we chose to estimate the energetic cost of a representative virus for this category, namely, the influenza virus. The influenza virus and T4 phage are functionally and evolutionarily very different viruses. Yet, surprisingly, they have a very similar per-virus cost, regardless of whether the total or the direct cost estimates are being considered. This is primarily due to the fact that they have a similar translational cost, which dominates all other costs. And, their comparable cost of translation is due to the fact that these viruses have similar dimensions and are both composed of about a million amino acids. Perhaps even more surprising is that both viral infections have very similar average power consumptions, on the order of 100 fW, despite their different durations.

Even with its higher burst size, an influenza infection has a total cost that is just 1% of the total cost of a eukaryotic cell. This is because a typical eukaryotic cell is estimated to have much higher energy supply than a typical bacterium under the same growth conditions. So far in our estimates, we do not account for the possible inefficiencies at various stages of the viral infection, which may drain more of the host energy than we estimated. Specifically, burst sizes are typically reported from plaque assays, which count the number of infectious virions that create plaques. However, we don't have a good estimate for the number of non-infectious viruses that arise from faulty genome replication, transcription, or viral assembly, for example. This point is especially important when considering RNA-based viruses such as influenza or HIV, which have higher mutation rates ($10^{-4}$-$10^{-6}$ mutations per base pair per generation; (32)) compared to dsDNA viruses such as T4 ($10^{-6}$-$10^{-8}$ mutations per base pair per generation; ((32)). As a result of these higher error rates, RNA-based viruses may have greater hidden costs associated with aborted viral synthesis or a greater fraction of faulty and non-infectious virions.

Second, even infectious viruses cannot all be guaranteed to enter the lytic cycle upon infecting a host cell. In support of these statements is the finding that only about 50% of PBCV-1 viral progeny are infectious (5). In fact, only 10% of influenza-infected host cells have been shown to generate infectious virions (33), demonstrating the cumulative inefficiency of an influenza infection. Hence, counting plaques to measure viral burst sizes may be analogous to making estimates of the human population by counting only individuals who have children. As such, single-cell studies of viral infection could provide a detailed breakdown of inefficiencies at various steps of the viral life-cycle and enable more exact cost estimates. We further explore other factors related to the fractional cost of influenza and T4 infections in the SI section XIV.

Finally, there is a great need for estimates of the effective population sizes of different viruses within their natural environments. With current effective population size estimates for viruses it appears that selection likely determines the fate of genetic elements for the majority of viruses, which have on average 28 nm radii (22) (Figure 5E, SI Table 4). However, for larger viruses (R > 200 nm), the diminishing, fractional cost of a gene may enable the interference of genetic drift to the extent that neutral genetic elements could persist in the viral population. The result of such a phenomenon could be genome expansions in the form of gene duplication events, cooption of previously noncoding, horizontally transferred elements into functional genes and regulatory domains, and perhaps, even a trend towards greater autonomy over large evolutionary time-scales. This effect may explain the unusual number of duplication events in the genomes of giant viruses such as that of the Mimivirus (34, 35). Perhaps this effect has also allowed enough genomic expansion and novelty for certain large viruses to jump the barrier between obligate entities and self-replicating organisms.

## Acknowledgements

We are grateful to David Baltimore, Forest Rohwer, Thierry Mora, Aleksandra Walczak, Ry Young, David Van Valen, Georgi Marinov, Elsa Birch, Yinon Bar-On and Ty Roach for their many insightful recommendations. This study was supported by the National Science Foundation Graduate Research Fellowship (grant no. DGE-1144469), The John Templeton Foundation (Boundaries of Life Initiative, grant ID 51250), the National Institute of Health's Maximizing Investigator's Research Award (grant no. RFA-GM-17-002), the National Institute of Health's Exceptional Unconventional Research Enabling Knowledge Acceleration (grant no. R01- GM098465), and the National Science Foundation (grant no. NSF PHY11-25915) through the 2015 Cellular Evolution course at the Kavli Institute for Theoretical Physics.

**Citations**

1. Kutter E & Sulakvelidze A (2004) *Bacteriophages: biology and applications* (CRC Press).
2. Hadas H, Einav M, Fishov I, & Zaritsky A (1997) Bacteriophage T4 development depends on the physiology of its host Escherichia coli. *Microbiology* 143(1):179-185.
3. Ahuka-Mundeke S*, et al.* (2010) Full-length genome sequence of a simian immunodeficiency virus (SIV) infecting a captive agile mangabey (Cercocebus agilis) is closely related to SIVrcm infecting wild red-capped mangabeys (Cercocebus torquatus) in Cameroon. *Journal of General Virology* 91(12):2959-2964.
4. Middelboe M (2000) Bacterial growth rate and marine virus–host dynamics. *Microbial Ecology* 40(2):114-124.
5. Van Etten JL, Burbank DE, Xia Y, & Meints RH (1983) Growth cycle of a virus, PBCV-1, that infects Chlorella-like algae. *Virology* 126(1):117-125.
6. Golec P, Karczewska-Golec J, Łoś M, & Węgrzyn G (2014) Bacteriophage T4 can produce progeny virions in extremely slowly growing Escherichia coli host: comparison of a mathematical model with the experimental data. *FEMS microbiology letters* 351(2):156-161.
7. Rosenwasser S, Ziv C, van Creveld SG, & Vardi A (2016) Virocell Metabolism: Metabolic Innovations During Host–Virus Interactions in the Ocean. *Trends in Microbiology* 24(10):821-832.
8. Lindell D*, et al.* (2007) Genome-wide expression dynamics of a marine virus and host reveal features of co-evolution. *Nature* 449(7158):83-86.
9. Thai M*, et al.* (2015) MYC-induced reprogramming of glutamine catabolism supports optimal virus replication. *Nature communications* 6.
10. Chang C-W, Li H-C, Hsu C-F, Chang C-Y, & Lo S-Y (2009) Increased ATP generation in the host cell is required for efficient vaccinia virus production. *Journal of biomedical science* 16(1):1.
11. Roux S*, et al.* (2016) Ecogenomics and potential biogeochemical impacts of globally abundant ocean viruses. *Nature*.
12. Kim H & Yin J (2004) Energy-efficient growth of phage Qβ in Escherichia coli. *Biotechnology and bioengineering* 88(2):148-156.
13. Birch EW, Ruggero NA, & Covert MW (2012) Determining host metabolic limitations on viral replication via integrated modeling and experimental perturbation. *PLoS Comput Biol* 8(10):e1002746.
14. Stray SJ & Air GM (2001) Apoptosis by influenza viruses correlates with efficiency of viral mRNA synthesis. *Virus research* 77(1):3-17.
15. Delbrück M (1945) The burst size distribution in the growth of bacterial viruses (bacteriophages). *Journal of bacteriology* 50(2):131.
16. Kaleta C, Schäuble S, Rinas U, & Schuster S (2013) Metabolic costs of amino acid and protein production in Escherichia coli. *Biotechnology journal* 8(9):1105-1114.
17. Phillips R, Kondev J, Theriot J, & Garcia H (2012) *Physical biology of the cell* (Garland Science).
18. Lynch M & Marinov GK (2015) The bioenergetic costs of a gene. *Proceedings of the National Academy of Sciences* 112(51):15690-15695.
19. Milo R, Jorgensen P, Moran U, Weber G, & Springer M (2010) BioNumbers—the database of key numbers in molecular and cell biology. *Nucleic acids research* 38(suppl 1):D750-D753.
20. Djamali E, Nulton JD, Turner PJ, Rohwer F, & Salamon P (2012) Heat output by marine microbial and viral communities.

21. Jover LF, Effler TC, Buchan A, Wilhelm SW, & Weitz JS (2014) The elemental composition of virus particles: implications for marine biogeochemical cycles. *Nature Reviews Microbiology* 12(7):519-528.
22. Brum JR, Schenck RO, & Sullivan MB (2013) Global morphological analysis of marine viruses shows minimal regional variation and dominance of non-tailed viruses. *The ISME journal* 7(9):1738-1751.
23. Wagner A (2005) Energy constraints on the evolution of gene expression. *Molecular biology and evolution* 22(6):1365-1374.
24. Koonin EV (2015) Energetics and population genetics at the root of eukaryotic cellular and genomic complexity. *Proceedings of the National Academy of Sciences* 112(52):15777-15778.
25. Charlesworth B (2009) Effective population size and patterns of molecular evolution and variation. *Nature Reviews Genetics* 10(3):195-205.
26. Novitsky V, Wang R, Lagakos S, & Essex M (2010) HIV-1 subtype C phylodynamics in the global epidemic. *Viruses* 2(1):33-54.
27. Bedford T, Cobey S, & Pascual M (2011) Strength and tempo of selection revealed in viral gene genealogies. *BMC Evolutionary Biology* 11(1):1.
28. Van Etten JL, Graves MV, Müller DG, Boland W, & Delaroque N (2002) Phycodnaviridae–large DNA algal viruses. *Archives of virology* 147(8):1479-1516.
29. Maynard ND, Gutschow MV, Birch EW, & Covert MW (2010) The virus as metabolic engineer. *Biotechnology journal* 5(7):686-694.
30. Anantharaman K*, et al.* (2014) Sulfur oxidation genes in diverse deep-sea viruses. *Science* 344(6185):757-760.
31. Roux A (2014) Reaching a consensus on the mechanism of dynamin? *F1000prime reports* 6.
32. Lauring AS, Frydman J, & Andino R (2013) The role of mutational robustness in RNA virus evolution. *Nature Reviews Microbiology* 11(5):327-336.
33. Brooke CB*, et al.* (2013) Most influenza a virions fail to express at least one essential viral protein. *Journal of virology* 87(6):3155-3162.
34. Raoult D*, et al.* (2004) The 1.2-megabase genome sequence of Mimivirus. *Science* 306(5700):1344-1350.
35. Suhre K (2005) Gene and genome duplication in Acanthamoeba polyphaga Mimivirus. *Journal of virology* 79(22):14095-14101.
36. Shepherd CM*, et al.* (2006) VIPERdb: a relational database for structural virology. *Nucleic acids research* 34(suppl 1):D386-D389.

**Table 1.** The direct, opportunity and total energetic costs of viral processes for T4 and influenza. The T4 infection costs are estimated based on an average burst size of 200, and the influenza infection costs are based on an average burst size of 6000. Direct costs shown represent the number of phosphate bonds directly hydrolyzed during the viral lifecycle (~P), whereas the total costs include both direct costs as well as opportunity costs (^P) incurred during the viral life-cycle (P). Empty cells correspond to viral processes that did not result in an energetic cost or were non-applicable to the given virus. See SI sections V-XI.

| | | | Replication | Transcription | Viral Entry | Packaging | Intracellular Transport | Viral Exit | Translation | Sum |
|---|---|---|---|---|---|---|---|---|---|---|
| direct cost (~P) | Per Virion | T4 | $4 \times 10^6$ | $7 \times 10^5$ | - | $3 \times 10^5$ | - | - | $7 \times 10^6$ | $10^7$ |
| | | flu | $3 \times 10^5$ | $7 \times 10^4$ | - | - | $10^3$ | $2 \times 10^6$ | $10^7$ | $10^7$ |
| | Per Infection | T4 | $9 \times 10^8$ | $10^8$ | - | $7 \times 10^7$ | - | - | $10^9$ | $3 \times 10^9$ |
| | | flu | $2 \times 10^9$ | $4 \times 10^8$ | $10^3$ | - | $6 \times 10^6$ | $10^{10}$ | $6 \times 10^{10}$ | $8 \times 10^{10}$ |
| opportunity cost (^P) | Per Virion | T4 | $10^7$ | $7 \times 10^5$ | - | - | - | - | $3 \times 10^7$ | $4 \times 10^7$ |
| | | flu | $8 \times 10^5$ | $2 \times 10^5$ | - | - | - | $3 \times 10^7$ | $5 \times 10^7$ | $9 \times 10^7$ |
| | Per Infection | T4 | $2 \times 10^9$ | $10^8$ | - | - | - | - | $6 \times 10^9$ | $8 \times 10^9$ |
| | | flu | $5 \times 10^9$ | $10^9$ | - | - | - | $2 \times 10^{11}$ | $3 \times 10^{11}$ | $5 \times 10^{11}$ |
| total cost (P) | Per Virion | T4 | $2 \times 10^7$ | $10^6$ | - | $3 \times 10^5$ | - | - | $4 \times 10^7$ | $6 \times 10^7$ |
| | | flu | $10^6$ | $3 \times 10^5$ | - | - | $10^3$ | $4 \times 10^7$ | $6 \times 10^7$ | $10^8$ |
| | Per Infection | T4 | $3 \times 10^9$ | $3 \times 10^8$ | - | $7 \times 10^7$ | - | - | $8 \times 10^9$ | $10^{10}$ |
| | | flu | $6 \times 10^9$ | $2 \times 10^9$ | $10^3$ | - | $6 \times 10^6$ | $2 \times 10^{11}$ | $4 \times 10^{11}$ | $6 \times 10^{11}$ |

**Figure 1.** The energetics of a T4 phage infection. The direct, opportunity and total costs of viral processes are denoted and can be distinguished by their units (~P, ^P and P, respectively). The energetic requirements of transcription (step 3), translation (step 4), genome replication (step 5), and genome packaging (step 7) are shown. See SI sections V-XI and Table 1. In this figure, when the opportunity cost component is left out, it can be assumed that it is equal to zero.

**Figure 2.** The energetics of an influenza infection. The direct, opportunity and total costs of viral processes are denoted and can be distinguished by their units (~P, ^P and P, respectively). The energetic requirements of viral entry (steps 2,3), intracellular transport (steps 4,5,9), transcription (step 6), translation (step 7), genome replication (step 8) and viral exit (step 10) are shown. See SI sections V-XI and Table 1. In this figure, when the opportunity cost component is left out, it can be assumed that it is equal to zero.

**Figure 3.** A breakdown of the direct cost (top) and the total cost (bottom) of various viral processes during T4 (left) and influenza (right) viral infections (normalized to the sum of all costs during an infection, as shown in the center of each pie chart). The direct cost of a T4 phage infection is approximately $3 \times 10^9$ ~P (top) while the total cost is $10^{10}$ P (bottom). The direct and total costs of an influenza infection are approximately ~$8 \times 10^{10}$ ~P and $6 \times 10^{11}$ P, respectively. Numbers are rounded to the nearest percent, and viral processes costing below 0.5% of the infection's cost are not shown. See SI sections V-XI for energetic cost estimates for viral entry, intracellular transport, transcription, viral assembly, and viral exit.

**Figure 4.** Generalizing viral energetics. A plot of the genome replication ($E_{REP}$) to translational cost ($E_{TL}$) ratio as a function of the virus inner radius, r. The plot uses the geometric parameters of viruses shown in SI Table 3, all of which are double-stranded DNA viruses with icosahedral geometries. The predicted numbers of amino acids and nucleotides are derived in SI Table 3.. Cost ratios are shown for both direct and total cost estimates. All viruses shown infect bacteria except Sputnik, which is a satellite virus of the giant Mimivirus. We have zoomed in on viruses Sputnik (r = 22 nm), P22 (r = 27.5 nm), T7 (r = 27.5 nm), HK97 (r = 30 nm), and Epsilon15 (r = 31.2 nm). The capsid structures for these representative viruses were obtained from the VIPERdb (36) and image sizes were scaled based on radii shown in SI Table 3 to accurately represent the relative sizes of each capsid. The critical radii for the total cost ($r_{crit-Tot}$) and the direct cost ($r_{crit-Dir}$) estimates are shown. We have also included the mean ($r_{mean}$= 25 nm) and standard deviation (gray vertical box, $\pm$6.5 nm) of viral capsid inner radii from 2,600 viruses collected by the Tara Oceans Expeditions (22). Note, here we have subtracted the mean capsid thickness (3 nm) from the mean capsid radius reported by Brum *et al*. to arrive at the mean *inner* capsid radius.

**Figure 5.** Evolutionary forces acting on genetic elements within viral genomes. A) Schematic of a virus as a spherical object, with an inner radius, r, an outer radius, R, and a capsid thickness, t. The capsid is composed of viral proteins, while the inner volume holds the viral genome. B) Positive and negative selective forces ($s_a$ and $s_c$) at a tug of war with the force of genetic drift, which scales as ${N_e}^{-1}$, where $N_e$ is the viral effective population size. C) A schematic of a genetic element within a viral genome. It is assumed to be non-functional ($s_a$ = 0) and non-transcribed, resulting in $|s_a|$ = $s_c$ = $E_g$, where $s_a$ corresponds to the net selection coefficient and $E_g$ corresponds to the fractional cost of a genetic element. D) The evolutionary forces acting on a genetic element within Virus A and Virus B genomes. The fractional cost of a genetic element in Virus B, $E_{g_VirusB}$, is 8 times higher than the fractional cost of the same element in Virus A, $E_{g_VirusA}$. Note, Virus A has twice the radius of Virus B, and therefore its genome is 8 times longer than that of Virus B (schematically represented by the number of genetic segments). Both viruses are assumed to have radii greater than critical radii, $r_{crit-Tot}$ and $r_{crit-Dir}$. E) $\text{Log}_{10}$ $E_g$ estimates for non-transcribed and neutral genetic elements of different lengths (1 – 10,000 base pairs) within the context of 30 dsDNA viruses ranging from ~20 nm to 400 nm in radius (SI Table 4; viruses with R > 50 nm are hypothetical dsDNA viruses). $\text{Log}_{10}$ $E_g$ estimates derived from both direct and total cost estimates are included (there is minimal difference between these estimates, which is not visible in this figure, see SI Table 4). Assuming $N_e$ = $10^5$, the region above the horizontal dashed line represents a selection-dominated regime, and the region below it represents a drift-dominated regime. For comparison, we have included the mean (vertical dashed line, 28 nm) and standard deviation (gray vertical box, $\pm$6.5 nm) of viral capsid radii obtained from 2600 viruses collected during the Tara Oceans Expeditions (22).

**Figure 1.**

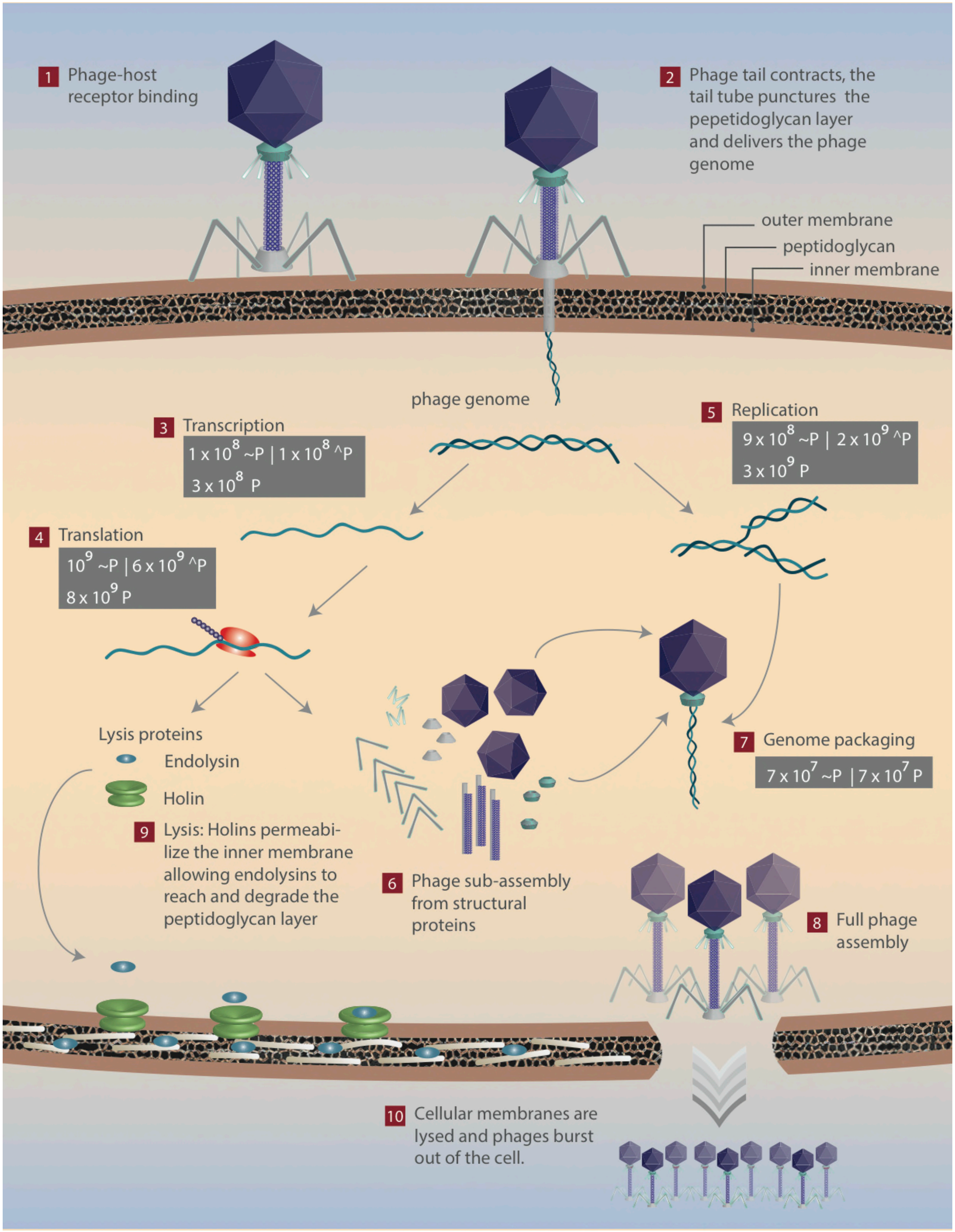

1 Phage-host receptor binding
2 Phage tail contracts, the tail tube punctures the pepetidoglycan layer and delivers the phage genome
outer membrane
peptidoglycan
inner membrane
phage genome
3 Transcription
1 x 10^8 ~P | 1 x 10^8 ^P
3 x 10^8 P
5 Replication
9 x 10^8 ~P | 2 x 10^9 ^P
3 x 10^9 P
4 Translation
10^9 ~P | 6 x 10^9 ^P
8 x 10^9 P
Lysis proteins
Endolysin
Holin
9 Lysis: Holins permeabi-lize the inner membrane allowing endolysins to reach and degrade the peptidoglycan layer
6 Phage sub-assembly from structural proteins
7 Genome packaging
7 x 10^7 ~P | 7 x 10^7 P
8 Full phage assembly
10 Cellular membranes are lysed and phages burst out of the cell.

**Figure 2.**

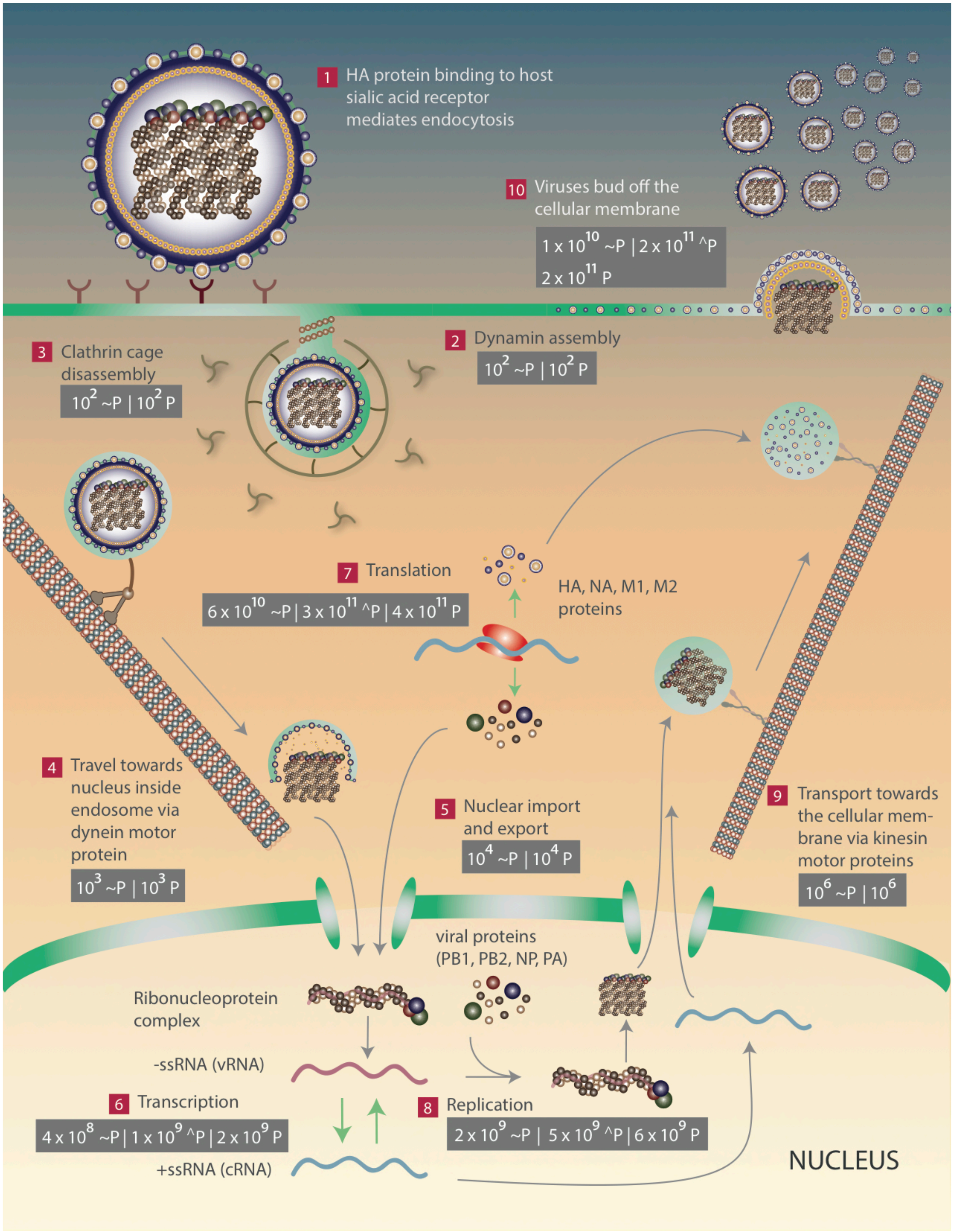

1 HA protein binding to host sialic acid receptor mediates endocytosis
10 Viruses bud off the cellular membrane
1 x 10^10 ~P | 2 x 10^11 ^P
2 x 10^11 P
2 Dynamin assembly
10^2 ~P | 10^2 P
3 Clathrin cage disassembly
10^2 ~P | 10^2 P
7 Translation
6 x 10^10 ~P | 3 x 10^11 ^P | 4 x 10^11 P
HA, NA, M1, M2 proteins
4 Travel towards nucleus inside endosome via dynein motor protein
10^3 ~P | 10^3 P
5 Nuclear import and export
10^4 ~P | 10^4 P
9 Transport towards the cellular membrane via kinesin motor proteins
10^6 ~P | 10^6
viral proteins (PB1, PB2, NP, PA)
Ribonucleoprotein complex
-ssRNA (vRNA)
6 Transcription
4 x 10^8 ~P | 1 x 10^9 ^P | 2 x 10^9 P
+ssRNA (cRNA)
8 Replication
2 x 10^9 ~P | 5 x 10^9 ^P | 6 x 10^9 P
NUCLEUS

**Figure 3.**

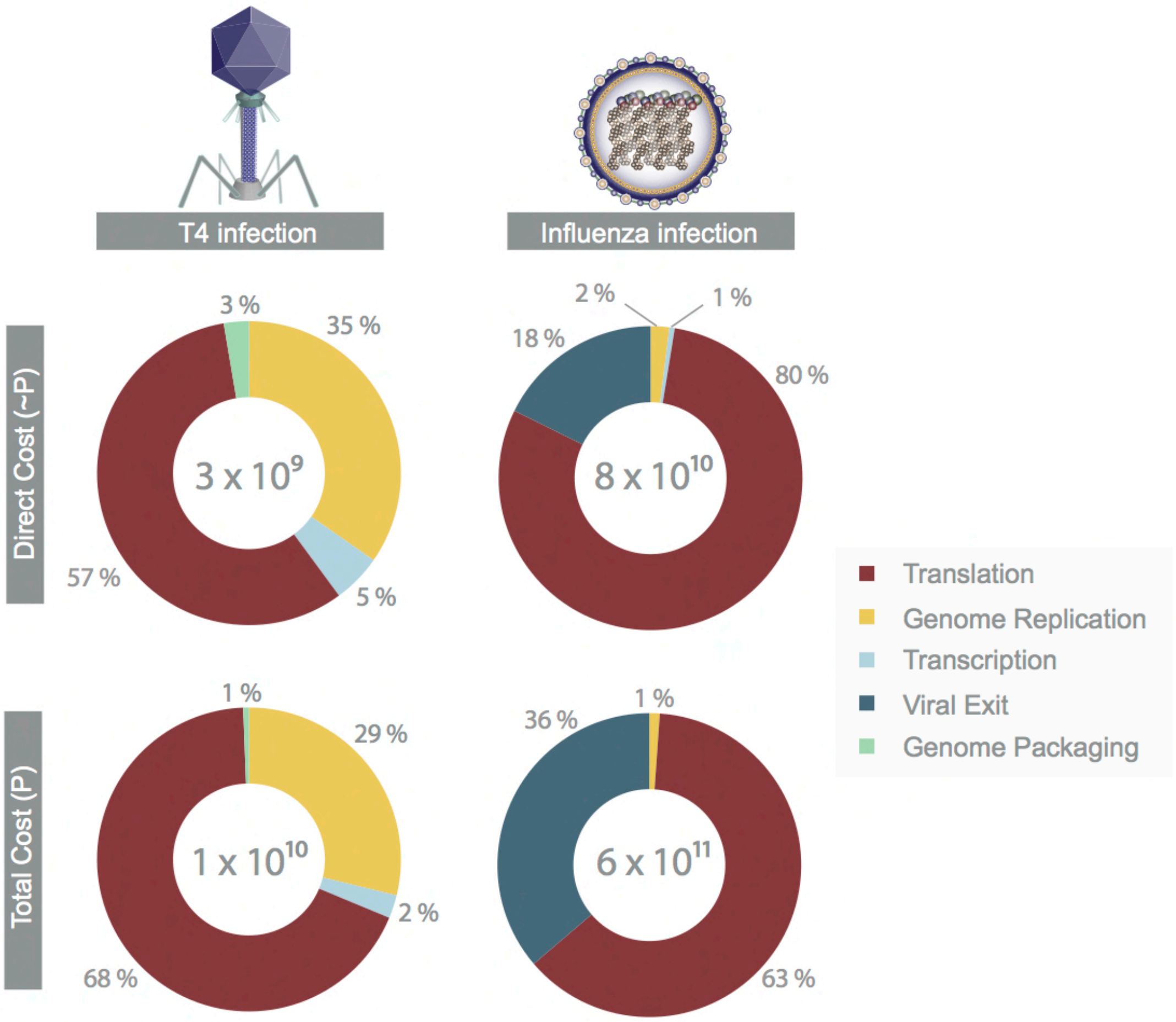

**Figure 4.**

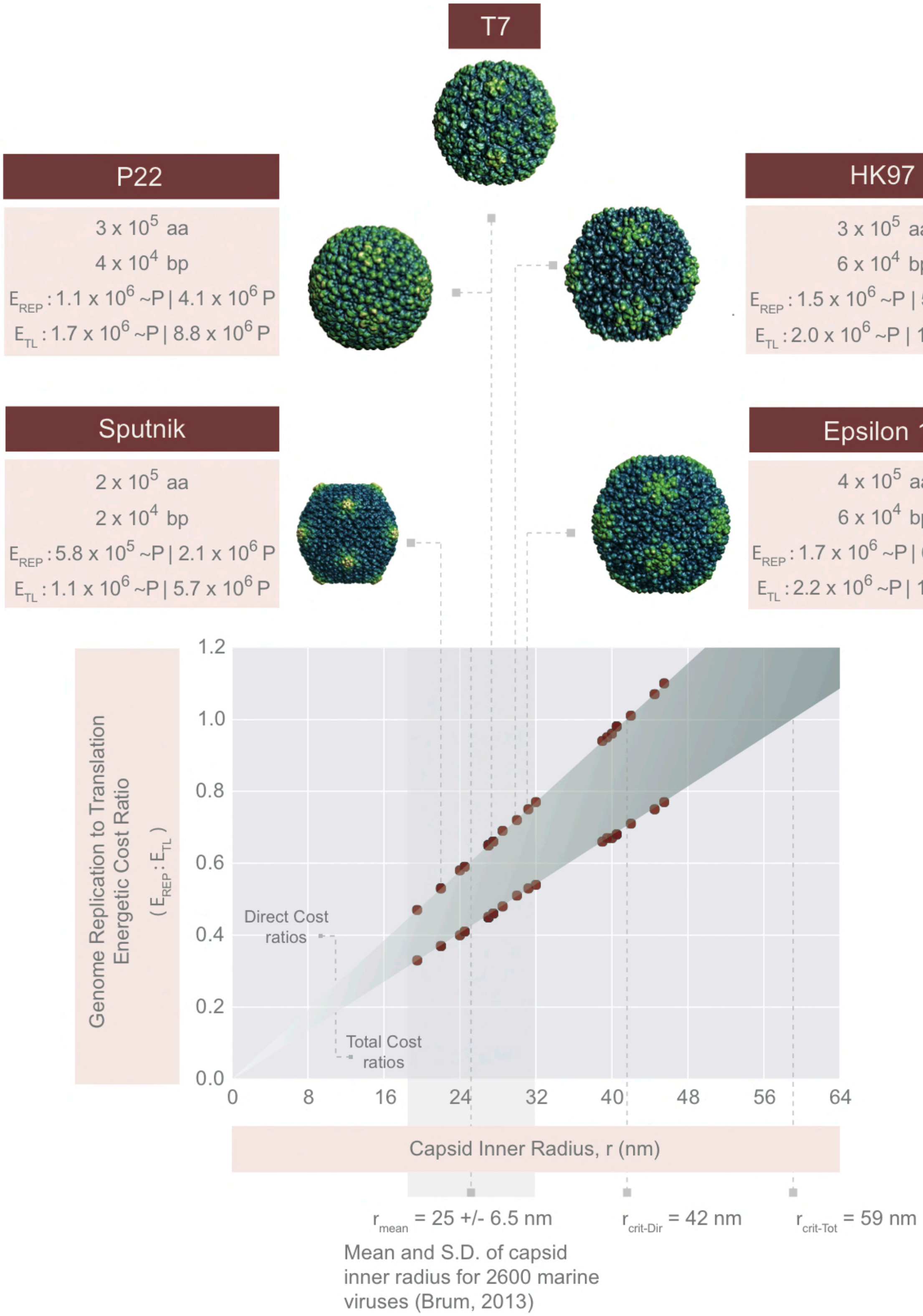
T7
P22
3 x 10^5 aa
4 x 10^4 bp
E_REP : 1.1 x 10^6 ~P | 4.1 x 10^6 P
E_TL : 1.7 x 10^6 ~P | 8.8 x 10^6 P
HK97
3 x 10^5 aa
6 x 10^4 bp
E_REP : 1.5 x 10^6 ~P | 5.3 x 10^6 P
E_TL : 2.0 x 10^6 ~P | 1.1 x 10^7 P
Sputnik
2 x 10^5 aa
2 x 10^4 bp
E_REP : 5.8 x 10^5 ~P | 2.1 x 10^6 P
E_TL : 1.1 x 10^6 ~P | 5.7 x 10^6 P
Epsilon 15
4 x 10^5 aa
6 x 10^4 bp
E_REP : 1.7 x 10^6 ~P | 6.0 x 10^6 P
E_TL : 2.2 x 10^6 ~P | 1.1 x 10^7 P
Genome Replication to Translation Energetic Cost Ratio
( E_REP : E_TL )
1.2
1.0
0.8
0.6
0.4
0.2
0.0
Direct Cost ratios
Total Cost ratios
0
8
16
24
32
40
48
56
64
Capsid Inner Radius, r (nm)
r_mean = 25 +/- 6.5 nm
r_crit-Dir = 42 nm
r_crit-Tot = 59 nm
Mean and S.D. of capsid inner radius for 2600 marine viruses (Brum, 2013)

**Figure 5.**

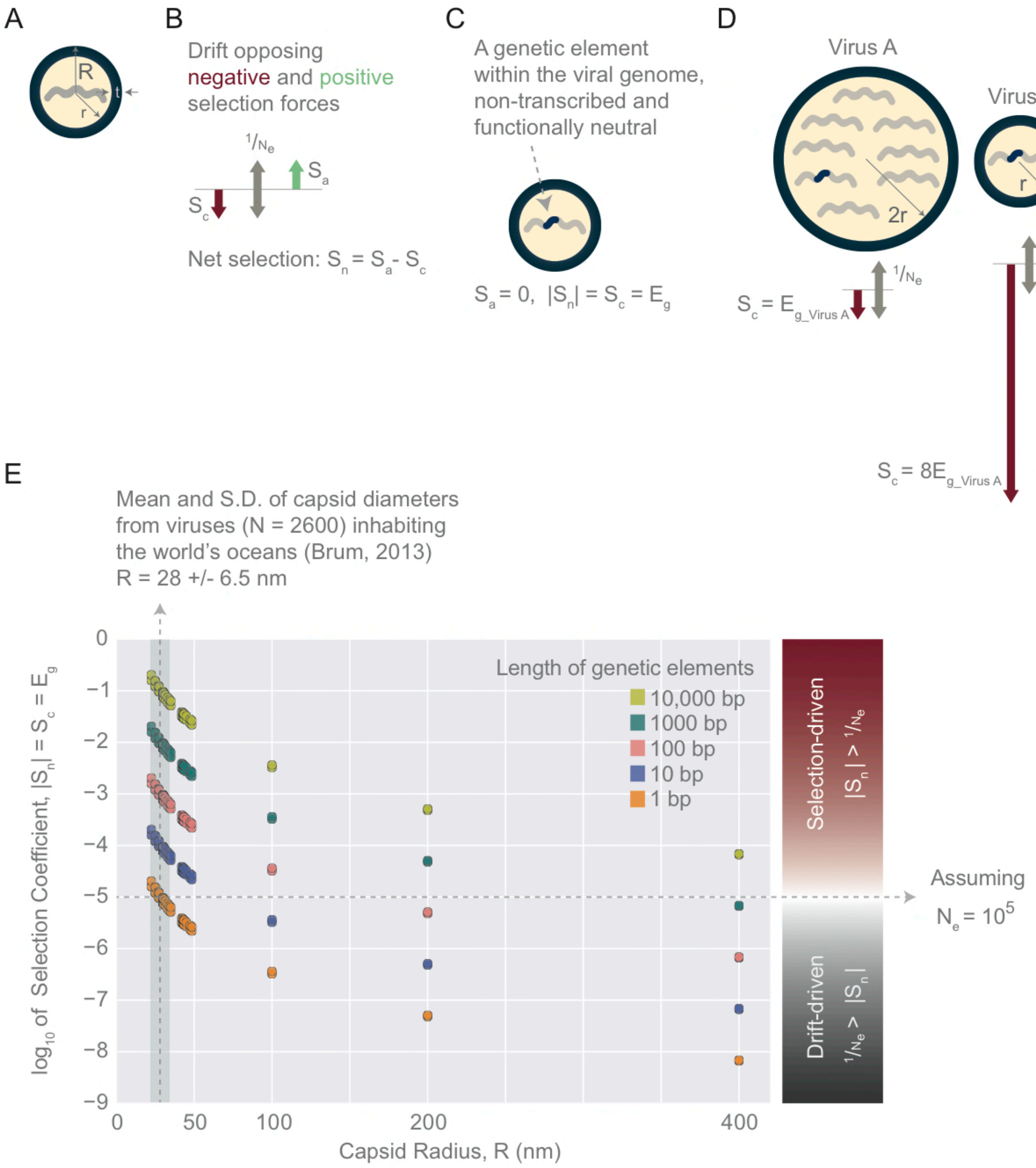

A
R
t
r
B
Drift opposing
negative and positive
selection forces
1/Ne
Sa
Sc
Net selection: Sn = Sa - Sc
C
A genetic element
within the viral genome,
non-transcribed and
functionally neutral
Sa = 0, |Sn| = Sc = Eg
D
Virus A
Virus B
2r
r
1/Ne
Sc = Eg_Virus A
Sc = 8Eg_Virus A
E
Mean and S.D. of capsid diameters
from viruses (N = 2600) inhabiting
the world's oceans (Brum, 2013)
R = 28 +/- 6.5 nm
Length of genetic elements
10,000 bp
1000 bp
100 bp
10 bp
1 bp
Selection-driven
|Sn| > 1/Ne
Drift-driven
1/Ne > |Sn|
Assuming
Ne = 10^5
log10 of Selection Coefficient, |Sn| = Sc = Eg
Capsid Radius, R (nm)
0
−1
−2
−3
−4
−5
−6
−7
−8
−9
0
50
100
200
400